\documentclass[11pt, prd, amsmath, nofootinbib]{revtex4-1} 
\usepackage[colorlinks, linkcolor=blue]{hyperref}
\usepackage{graphicx} 
\usepackage{subcaption}
\usepackage{amsmath}
\usepackage{array} 
\usepackage[dvipsnames]{xcolor}

\begin{document}

\title{Instability of Holographic Superfluids in Optical Lattice}
\author{Peng Yang$^1$}\email{yangpeng18@mails.ucas.edu.cn}
\author{Xin Li$^{1,2}$}\email{xin.z.li@helsinki.fi}
\author{Yu Tian$^{1,3}$}\email{ytian@ucas.ac.cn}

\affiliation{$^1$School of Physical Sciences, University of Chinese Academy of Sciences, Beijing 100049, China}
\affiliation{$^2$Department of Physics P.O. Box 64, FI-00014 University of Helsinki, Finland}
\affiliation{$^3$Institute of Theoretical Physics, Chinese Academy of Sciences, Beijing 100190, China}

\begin{abstract}
The instability of superfluids in optical lattice has been investigated using the holographic model. The static and steady flow solutions are numerically obtained from the static equations of motion and the solutions are described as Bloch waves with different Bloch wave vector $k$. Based on these Bloch waves, the instability is investigated at two levels. At the linear perturbation level, we show that there is a critical $k_{c}$ above which the superflow is unstable. At the fully nonlinear level, the intermediate state and final state of unstable superflow are identified through numerical simulation of the full equations of motion. The results show that during the time evolution, the unstable superflow will undergo a chaotic state \textcolor{black}{with soliton generation}. The system will settle down to a stable state with $k<k_{c}$ eventually, with a smaller current and a larger condensate.
\end{abstract}

\maketitle

\tableofcontents

\section{Introduction}

As an artificial lattice structure, optical lattice, has a close physical resemblance to the periodic coulomb potential that felt by electrons in solid crystal. While it is not like electrons moving between positive ions in a nanometer dimension, optical lattice can be manipulated with a typical dimension much larger than the conventional crystal—micrometer dimension. Due to this virtue, optical lattice has provided a broad platform for experimental and theoretical physicists to explore \textcolor{black}{much richer and more} fundamental properties in condensed matter physics, such as instability \cite{ PhysRevA.64.061603,Wu_2003}, Landau-Zener tunneling \cite{1996PhRvL..76.4504N,2000PhRvA..61b3402W,2002PhRvA..65f3612C,2003PhRvL..91w0406J,Wu_2003,2020PhRvL.125u3401G}, superfluid-Mott insulator quantum phase transition \cite{1998PhRvL..81.3108J,2002PhRvA..66c1601K, 2002PhRvL..89q0402S, 2004PhRvA..70d3612C, 2005PhRvA..71d3601Z,2006PhRvA..73b3610C} and so on. Additionally, by controlling the frequency of the lasers \textcolor{black}{in the course of time}, the time-dependent optical lattice can be used for driving the system to explore the floquet dynamics \cite{2015AdPhy..64..139B,PhysRevLett.115.205301,PhysRevLett.121.233603,PhysRevX.9.011047,2019NatPh..15.1168S}, which are getting more interesting recently.

The instability (or stability) of superfluids is always one of the most fundamental and important properties and usually many physical phenomena are related with it. Since the optical lattice starts to be used for studying cold atom physics, many kinds of research works about the instability \textcolor{black}{have been carried} out in various respects. There are Landau or dynamical instability \cite{2001PhRvE..64e6615B,2001PhRvE..63c6612B,PhysRevA.64.061603,2003NJPh....5...71C,Wu_2003,2004PhRvL..93n0406F,2004PhRvA..70d3625M,PhysRevLett.95.180403,PhysRevLett.95.110403,2006PhRvA..74e3611I,2006JPhB...39S.101K,2007PhRvA..75c3621H,ZHANG20081147,2009ChPhL..26a6701L,2014JPSJ...83l4001A,PhysRevA.93.013628,PhysRevLett.86.4447,Cataliotti_2003,PhysRevLett.93.140406, PhysRevA.72.013603,Konabe_2006}, parametric instabilities \cite{2004JPhB...37S.257R,2005PhRvA..71f1602K,2005PhRvL..95q0404G,2017PhRvX...7b1015L,PhysRevX.10.011030}, modulational instability \cite{2002PhRvA..65b1602K,2004PhRvE..69a7601R,2006PhRvL..96o0402B,PhysRevA.87.013633,PhysRevE.89.052917} and low-acceleration instability \cite{PhysRevLett.93.230401}, etc. In this paper we will focus on the Landau and dynamical instability only and simultaneously.

Since the first theoretical study about the Landau and dynamical instability for superfluids in optical lattice \cite{PhysRevA.64.061603} using the \textcolor{black}{Gross-Pitaevskii (GP)} equation, this topic has been studied extensively. \cite{2001PhRvE..63c6612B,2001PhRvE..64e6615B} explore the cases both for repulsive and attractive atomic interactions, \cite{ZHANG20081147} adds the influence of three-body interactions. \cite{2009ChPhL..26a6701L} shows that the states in excited bands always have Landau instability and \cite{2004PhRvA..70d3625M} investigates the effect of transverse excitations for higher dimensional systems. \cite{PhysRevLett.95.180403} reveals that the dynamical instability is the origin of the spatial domain formation in spin-1 atomic condensates. All these research works relied on the mean-field theory, i.e., GP equation, while in addition to this equation the Bose-Hubbard model is also used in \cite{2014JPSJ...83l4001A,PhysRevA.93.013628} that is beyond the mean-field theory.

\textcolor{black}{In the above research works, all the superfluid systems have no dissipation.} From the instability diagram \textcolor{black}{(Fig.5 in \cite{PhysRevA.64.061603}) we} can see that there are two distinct regimes about the instability of superflow. The first is the small optical lattice height regime and in this regime the dynamical instability regime is so narrow that the Landau instability can be studied solely; the second is the large optical lattice height regime where the dynamical instability regime spreads out and it can be detected in experiments, obviously. Experiment \cite{PhysRevLett.86.4447} shows that the regime where the dissipative process occurs to break down the superfluid phase agrees with the theoretical prediction about the regime of onset of Landau instability for a superfluid in shallow optical lattice; Experiment \cite{Cataliotti_2003} finds that for deep optical lattice there will be transition from a superfluid into an insulator when the superfluid is under dynamical instability. After that, experiments \cite{PhysRevLett.93.140406, PhysRevA.72.013603} both show that in the presence of thermal component the dissipative process can be related with the occurrence of Landau instability. Since the \textcolor{black}{original GP equation cannot apply to dissipative processes, until \cite{Konabe_2006} based on the modified GP equation with dissipation, the instability of superfluid in optical lattice with thermal component has not been} studied theoretically. Its result dramatically shows that when the thermal component is added the dynamical instability will \textcolor{black}{occur throughout the} Landau instability regime predicted from the original GP equation, which means the Landau instability and dynamical instability will always appear at the same time when there exists dissipation. So it is no longer suitable to investigate Landau instability or dynamical instability separately.

In this paper, we will use \textcolor{black}{the simplest} holographic superfluid model \cite{PhysRevLett.101.031601,Hartnoll_2008} to re-explore the Landau and dynamical instability
of superfluids in optical lattice simultaneously based on the advantage that such a model contains \textcolor{black}{superfluid and normal fluid (as thermal component with a finite temperature)} intrinsically, which \textcolor{black}{introduces the dissipation naturally and consistently} \textcolor{black}{\cite{Adams:2012pj}. In contrast, the dissipation in the modified GP equation is purely put by hand, without any relation to thermodynamics. In holography, the unstable region can be calculated by quasi-normal mode (QNM) analysis} and \textcolor{black}{the results show that there is a critical Bloch wave vector $k_{c}$ above which the Bloch states are unstable due to the Landau instability and dynamical instability. With the help of the calculation of sound speed we confirm that the Landau instability and dynamical instability also appear at the same time in the presence of optical lattice just like the homogenous case in \cite{2014JHEP...02..063A,2020arXiv201006232L}.}

\textcolor{black}{Beyond the linear perturbation analysis, the unstable superflow are also studied by the numerical simulation. In this part we use the unstable superflow state as the initial state for the evolution equation and evolve it for a long time with some small perturbation given at early time. The unstable modes of perturbation will grow exponentially at first and it will lead the whole system into a chaotic state, in which the solitons will appear and disappear (for higher dimension there will be vortices and the system can be considered as under a transient turbulence state \cite{2020arXiv201006232L}). Finally, along with some dissipative processes the system will reduce its current to become stable with the final wave vector $k<k_{c}$. These results can be tested in experiments.}

\textcolor{black}{This paper is organized as follows. In Sec.~\ref{Holographic Model} we introduce the holographic superfluid model that we use and \textcolor{black}{show how} the optical lattice is added. Then in Sec.~\ref{Bloch waves} \textcolor{black}{by solving the corresponding equations of motion} we get the static and steady-flow suferfluid states that expressed as Bloch waves with different \textcolor{black}{wave vector} $k$. The QNM analysis and the calculation of sound speed are in Sec.~\ref{Landau instability}. And in Sec.~\ref{Real time evolution} we present the dynamic processes of the evolution of an unstable superflow with numerical simulation. Finally, in Sec.~\ref{Summary} we give a summary.}

\section{Holographic Model}\label{Holographic Model}

The simplest holographic model to describe superfluids is given in \cite{PhysRevLett.101.031601}, where a complex scalar field $\Psi$ is coupled to a $U(1)$ gauge field $A_M$ in the $(3+1)$D gravity with a cosmological constant related to the AdS radius as $\Lambda=-3/{L^2}$. The action is
\begin{equation}
S=\frac{1}{16 \pi G_{4}} \int d^{4} x \sqrt{-g}\left(R+\frac{6}{L^{2}}+\frac{1}{e^{2}} \mathcal{L}_{M}\right),
\end{equation}
where $G_{4}$ is the gravitational constant in four dimensional spacetime. The first two terms in the parenthesis are the gravitational part of the Lagrangian with the Ricci scalar $R$ and the AdS radius $L$, and the third consists of all matter fields:
\begin{equation}
\mathcal{L}_{M}=-\frac{1}{4} F_{\mu \nu}F^{\mu \nu}-\left|D_{\mu} \Psi\right|^{2}-m^{2}|\Psi|^{2},
\end{equation}
where $D_{\mu}=\partial_{\mu} \Psi-i A_{\mu} \Psi$.

Since the backreaction of matter fields onto the spacetime geometry is not necessary for our problem, taking the probe limit ${e}\to\infty$ is a suitable and convenient choice, which means that we fix the background spacetime as the standard \textcolor{black}{Schwarzschild-AdS black brane} with metric
\begin{equation}
ds^{2}=\frac{L^{2}}{z^{2}}\left(-f(z)dt^{2}+\frac{1}{f(z)}dz^{2}+dx^{2}+dy^{2}\right),\qquad f(z)=1-\frac{z^{3}}{z_{H}^{3}}.
\end{equation}
\textcolor{black}{Here $z_H$ is the radius of the black brane horizon.} The Hawking temperature of this black brane is $T=\frac{3}{4\pi z_{H}}$, which is also the temperature of the boundary system by the holographic dictionary. Furthermore, due to the existence of the bulk black hole the boundary field theory will intrinsically have dissipation, since there will be energy flow absorbed into the horizon \cite{Adams:2012pj,Li:2019swh,Tian:2019fax} \textcolor{black}{during dynamic processes}. For numerical simplicity, we set $L=1=z_{H}$.

The equations of motion for $A_{\mu}$ and $\Psi$ are
\begin{equation}
\frac{1}{\sqrt{-g}} D_{\mu}\left(\sqrt{-g} D^{\mu} \Psi\right)-m^{2} \Psi=0,
\end{equation}
\begin{equation}
\frac{1}{\sqrt{-g}} \partial_{\mu}\left(\sqrt{-g} F^{\mu \nu}\right)=i\left(\Psi^{*} D^{\nu} \Psi-\Psi\left(D^{\nu} \Psi\right)^{*}\right).
\end{equation}
One can easily see that there is a trivial solution with $\Psi=0$. As we increase the chemical potential $\mu$, there will be a critical chemical potential $\mu_{c}$ above which a solution with nonzero $\Psi$ appears, indicating the break of $U(1)$ symmetry and the formation of superfluid condensate. Hereafter, we will focus on the case with $\mu=4.5>\mu_{c}$, i.e., the superfluid phase of the boundary field theory.

\subsection{Equations of motion and optical lattice}

Since the considered problem is based on putting superflow onto the optical lattice, the density of the superflow will be periodically modulated by the periodic potential (chemical potential plus external potential). For simplicity and without loss of generality, we take the axial gauge $A_{z}=0$ as usual and consider our bulk fields as functions of $(t,z,x)$ with the assumption that the direction of the optical lattice is along $x$. \textcolor{black}{It turns out that $A_{y}$ can be turned off in this case, so} the remaining fields in the bulk are $A_{t}\left(t,z,x\right)$, $A_{x}\left(t,z,x\right)$ and $\Psi\left(t,z,x\right)$. With all these simplifications the equations of motion become
\begin{equation}
z^{2}\left(-\frac{1}{f}\left(\partial_{t}^{2} A_{x}-\partial_{t} \partial_{x} A_{t}\right)+\partial_{z}\left(f \partial_{z} A_{x}\right)\right)=i\left(\Psi^{*} \partial_{x} \Psi-\Psi \partial_{x} \Psi^{*}-2 i \Psi A_{x} \Psi^{*}\right),
\end{equation}
\begin{equation}
-z^{2}\left(\partial_{t} \partial_{z} A_{t}+f \partial_{x} \partial_{z} A_{x}\right)=i f\left(\Psi^{*} \partial_{z} \Psi-\Psi \partial_{z} \Psi^{*}\right),
\end{equation}
\begin{equation}
z^{2}\left(-\partial_{t} \partial_{x} A_{t}+f \partial_{z}^{2}A_{t}+\partial_{x}^{2} A_{t}\right)=i\left(\Psi^{*} \partial_{t} \Psi-\Psi \partial_{t} \Psi^{*}\right)+2 A_{t} \Psi^{*} \Psi,
\end{equation}
\begin{equation}
-\frac{z^{2}}{f}\left(\partial_{t}-i A_{t}\right)^{2}\Psi+z^{4} \partial_{z}\left(z^{-2} f\right) \partial_{z} \Psi+z^{2} f \partial_{z}^{2} \Psi+z^{2}\left(\partial_{x}-i A_{x}\right)^{2} \Psi-m^{2} \Psi=0,
\end{equation}
and the corresponding static equations are
\begin{equation}\label{static Jx}
z^{2}\left(\partial_{z} f \partial_{z} A_{x}+f \partial_{z}^{2} A_{x}\right)=i\left(\Psi^{*} \partial_{x} \Psi-\Psi \partial_{x} \Psi^{*}-2 i \Psi A_{x} \Psi^{*}\right),
\end{equation}
\begin{equation}\label{static Jz}
-z^{2} \partial_{x} \partial_{z} A_{x}=i\left(\Psi^{*} \partial_{z} \Psi-\Psi\partial_{z} \Psi^{*}\right),
\end{equation}
\begin{equation}\label{static Jt}
z^{2}\left(f \partial_{z}^{2} A_{t}+\partial_{x}^{2} A_{t}\right)=2 A_{t} \Psi^{*} \Psi,
\end{equation}
\begin{equation}\label{static Psi}
\frac{z^{2}}{f} A_{t}^{2} \Psi+z^{4} \partial_{z}\left(z^{-2}f\right) \partial_{z} \Psi+z^{2} f \partial_{z}^{2} \Psi+z^{2}\left(\partial_{x}-i A_{x}\right)^{2} \Psi-m^{2} \Psi=0.
\end{equation}
Actually these equations (even the static ones) are too complicated to have nontrivial analytical solutions, so numerical calculation is needed.

From the asymptotic analysis of all these static fields we know
\begin{equation}
\Psi \sim z^{d-\Delta_{+}} \Psi_{+}+z^{d-\Delta_{-}} \Psi_{-}+\ldots,
\end{equation}
\begin{equation}\label{At}
A_{t} \sim a_{t}-z^{d-2} \rho+\ldots,
\end{equation}
\begin{equation}
A_{x} \sim a_{x}-z^{d-2} j_{x}+\ldots,
\end{equation}
where $\Delta_{\pm}=\frac{d \pm \sqrt{d^{2}+4 m^{2}}}{2}$ with $d=3$ in our case. The holographic dictionary tells us that with one of $\Psi_{\pm}$ being the source the other will be the related response, $a_{t}$ is the total potential, $\rho$ the \textcolor{black}{(conserved)} particle number density and $a_{x}$ the source of the \textcolor{black}{particle current density} $j_{x}$ in the boundary field theory. The optical lattice is then introduced by \textcolor{black}{including a potential $V(x)$ in} $a_{t}$, i.e. $a_{t}=\mu+V(x)$. It is convenient to choose $m^{2}=-2$ to make all calculations easier, in which case
\begin{equation}
	\Psi \sim z \Psi_{+}+z^{2} \Psi_{-}+O\left(z^{3}\right)=:z \psi,
\end{equation}
and we choose $\Psi_{+}$ as source and $\Psi_{-}$ as the response. Here we introduce the bulk field $\psi$ for numerical simplicity.
\begin{figure*}
\begin{subfigure}[b]{0.32\textwidth}
\includegraphics[width = \textwidth]{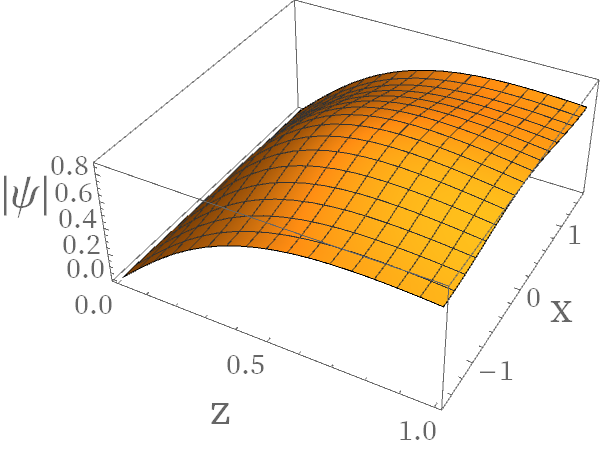}
\caption{amplitude of $\psi, k=0$}
\label{figblochwavek0}
\end{subfigure}
\begin{subfigure}[b]{0.32\textwidth}
\includegraphics[width = \textwidth]{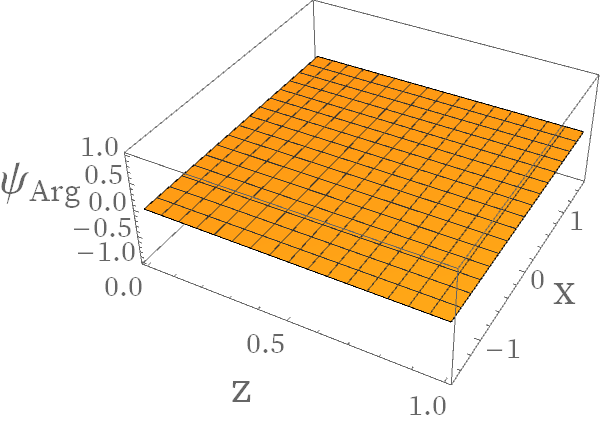}
\caption{angle of $\psi, k=0$}
\label{figwaveRIk0}
\end{subfigure}
\begin{subfigure}[b]{0.32\textwidth}
\includegraphics[width = \textwidth]{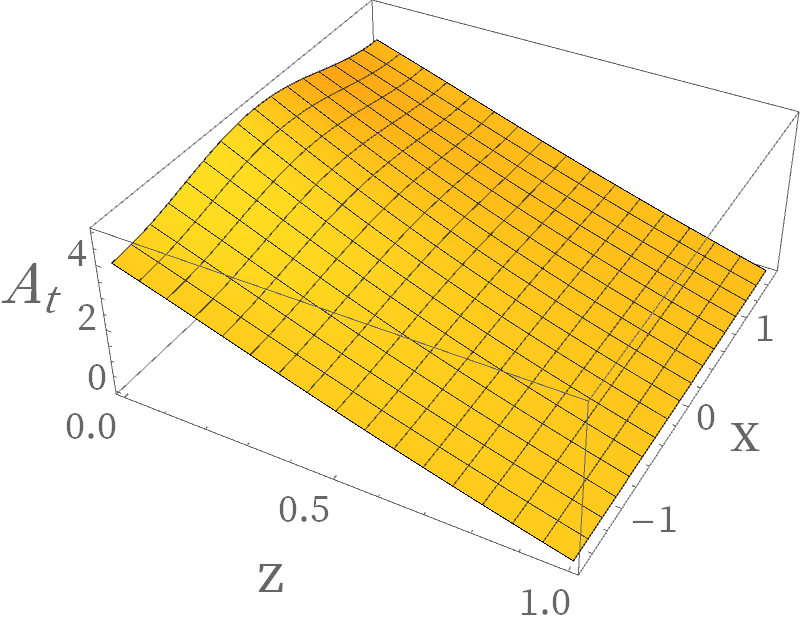}
\caption{$A_{t}, k=0$}
\label{figuvk0}
\end{subfigure}\\
\begin{subfigure}[b]{0.32\textwidth}
\includegraphics[width = \textwidth]{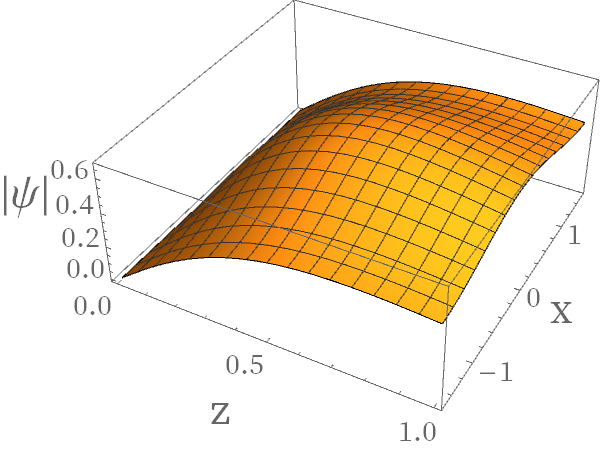}
\caption{amplitude of $\psi, k=0.75$}
\label{figblochwavek025}
\end{subfigure}
\begin{subfigure}[b]{0.32\textwidth}
\includegraphics[width = \textwidth]{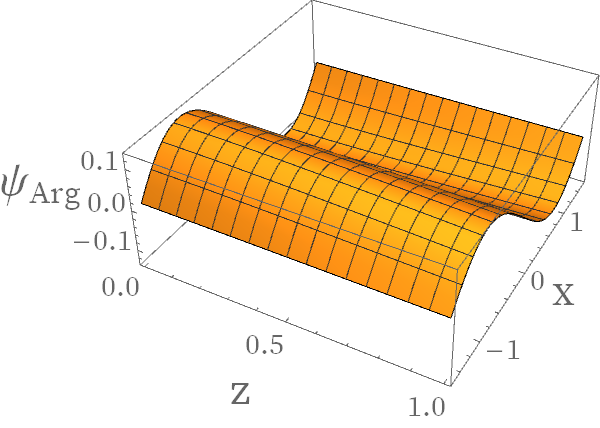}
\caption{angle of $\psi, k=0.75$}
\label{figwaveRIk025}
\end{subfigure}
\begin{subfigure}[b]{0.32\textwidth}
\includegraphics[width = \textwidth]{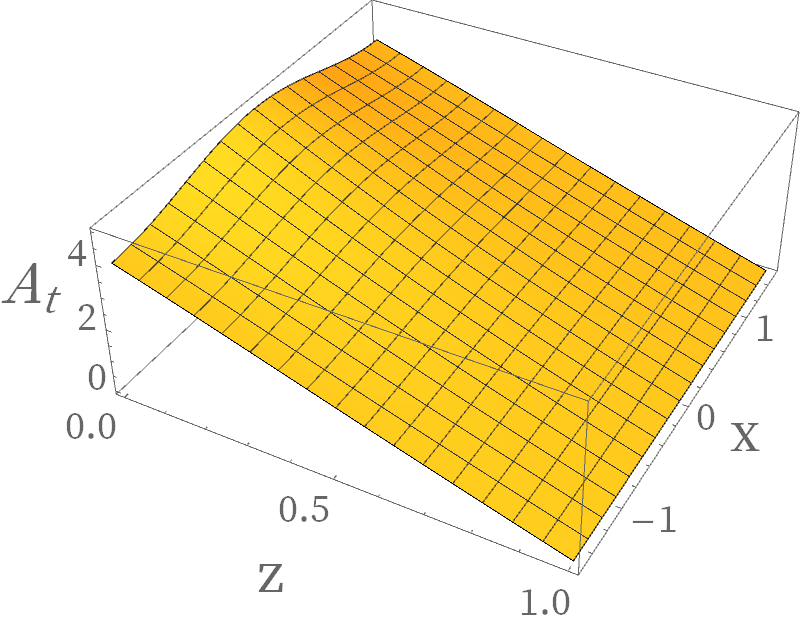}
\caption{$A_{t}, k=0.75$}
\label{figuvk025}
\end{subfigure}
\caption{The amplitude and \textcolor{black}{phase angle} of the bulk field $\psi$ are plotted in subfigures \ref{figblochwavek0}, \ref{figwaveRIk0}, \ref{figblochwavek025} and \ref{figwaveRIk025}; The bulk field $A_{t}$ is plotted in subfigures \ref{figuvk0} and \ref{figuvk025}. All the fields are calculated at $k=0$ and $0.75$.}
\label{figblochband}
\end{figure*}
\begin{figure*}
\begin{subfigure}[b]{0.48\textwidth}
\includegraphics[width =\textwidth]{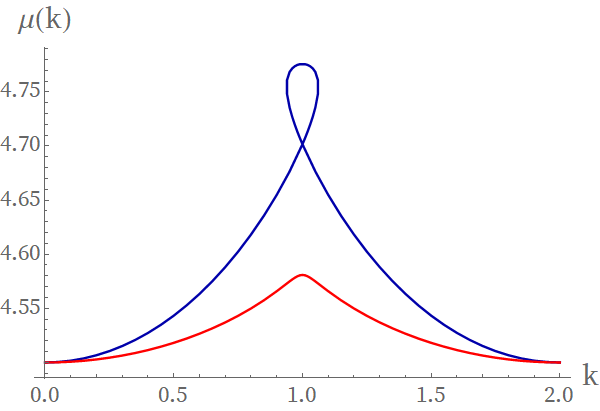}
\caption{Lowest Bloch band $\mu({k})$}
\label{fig bloch band}
\end{subfigure}
\begin{subfigure}[b]{0.48\textwidth}
\includegraphics[width =\textwidth]{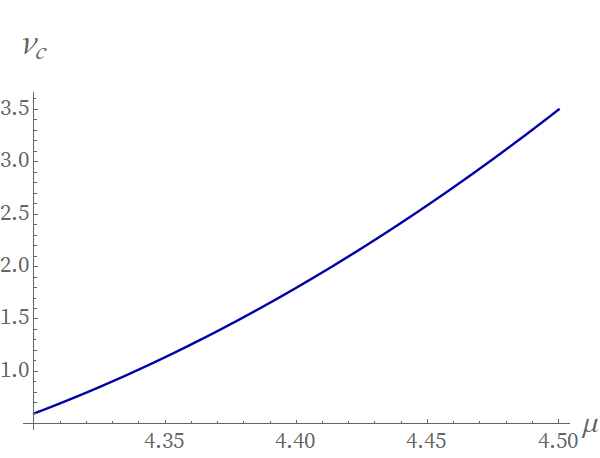}
\caption{Critical optical lattice strength $v_{c}$}
\label{fig bloch band2}
\end{subfigure}
\caption{\textcolor{black}{Fig.~\ref{fig bloch band} shows the lowest Bloch band $\mu(k)$ as a function of Bloch wave vector $k$ with fixed total particle number $N$, whose value is given by the case of chemical potential $\mu=4.5$, lattice strength $v=0.5$ (blue), 4 (red) and Bloch wave vector $k=0$. Fig.~\ref{fig bloch band2} shows the relation between critical optical lattice strength $v_{c}$ and chemical potential $\mu$. The Bloch band will \textcolor{black}{have loop structure when $v<v_{c}$.}}}
\label{figomegaq}
\end {figure*}

\section{Bloch waves and Bloch band in optical lattice}\label{Bloch waves}

From the Bloch theorem and with the source $\Psi_{+}$ turning off, we can use Bloch waves to describe the static response
\begin{equation}
\Psi_{-}\left(z,x\right)=\psi_{B}\left(z,x\right)e^{ikx}.
\end{equation}
Here, $\psi_{B}\left(z,x\right)$ is the Bloch wave \textcolor{black}{along the} $x$ direction, which has the same period as the optical lattice, i.e. $\psi_{B}\left(z,x+l\right)=\psi_{B}\left(z,x\right)$, \textcolor{black}{$k$ is the Bloch wave vector and $l$ the lattice constant (hereafter we choose $l=\pi$)}. Due to the fact that Bloch waves are complex functions, we can separate them into real and imaginary parts:
\begin{equation}\label{ansatz psi}
\Psi\left(z,x\right)=z\left[\psi_{R}\left(z,x\right)+i\psi_{I}\left(z,x\right)\right]e^{ikx}.
\end{equation}
Substituting the above equation into the static equations of motion \eqref{static Jx}-\eqref{static Psi}, we get the following five differential equations:
\begin{equation}\label{Jx}
\partial_{z} f \partial_{z} A_{x}+f \partial_{z}^{2} A_{x}=-2\left(\left(\psi_{R} \partial_{x} \psi_{I}-\psi_{I} \partial_{x} \psi_{R}\right)+\left(k-A_{x}\right)\left(\psi_{R}^{2}+\psi_{I}^{2}\right)\right),
\end{equation}
\begin{equation}\label{Jz}
\partial_{x} \partial_{z} A_{x}=2\left(\psi_{R} \partial_{z} \psi_{I}-\psi_{I} \partial_{z} \psi_{R}\right),
\end{equation}
\begin{equation}\label{Jt}
f \partial_{z}^{2}A_{t}+\partial_{x}^{2}A_{t}=2 A_{t}\left(\psi_{R}^{2}+\psi_{I}^{2}\right),
\end{equation}
\begin{equation}\label{Psi a}
\left(\frac{1}{f}A_{t}^{2}-z-\left(A_{x}-k\right)^{2}\right) \psi_{R}+\partial_{z}\left(f \partial_{z} \psi_{R}\right)+\left(\partial_{x}^{2} \psi_{R}-2\left(k-A_{x}\right) \partial_{x} \psi_{I}+\psi_{I} \partial_{x} A_{x}\right)=0,
\end{equation}
\begin{equation}\label{Psi b}
\left(\frac{1}{f} A_{t}^{2}-z-\left(A_{x}-k\right)^{2}\right) \psi_{I}+\partial_{z}\left(f \partial_{z} \psi_{I}\right)+\left(\partial_{x}^{2} \psi_{I}+2\left(k-A_{x}\right) \partial_{x} \psi_{R}-\psi_{R} \partial_{x} A_{x}\right)=0.
\end{equation}
\textcolor{black}{Among them, Eq.\eqref{Jz} will be satisfied automatically if the other four equations are solved.} To obtain the static or steady-flow states of the superfluid, we impose the source free boundary conditions for $\Psi_{+}$ and $A_{x}$ at the conformal boundary, regular conditions for $\Psi_{-}$ and $A_{t}=0$ at the horizon and periodic boundary conditions for $A_{t,x}$ and $\Psi$ in the $x$ direction. And the optical lattice structure is imposed by choosing $V(x)=v\cos(2x)$, i.e. $\left.A_{t}\right|_{z=0}=4.5+v\cos(2x)$, \textcolor{black}{with $v$ the height (or strength) of the optical lattice}. We solve these four static equations of motion \eqref{Jx}, \eqref{Jt}, \eqref{Psi a} and \eqref{Psi b} numerically with the Newton-Raphson method.

Fig.\ref{figblochband} shows the numerical solutions of the bulk fields $\psi$ and $A_{t}$, respectively, with two different Bloch wave vectors. Since $\psi$ is complex, we plot its amplitude and \textcolor{black}{phase angle} separately. The inexistence of node of \textcolor{black}{order parameter} means that these Bloch waves sit in the lowest Bloch band. In Fig.~\ref{fig bloch band}, we plot the Bloch band \textcolor{black}{in our holographic model}. To obtain it, we fix the total particle number, defined as
\begin{equation}
    N\equiv\int_{-\frac{\pi}{2}}^{\frac{\pi}{2}}dx\rho,
\end{equation}
and solve the static equations \eqref{Jx} and \eqref{Jt}-\eqref{Psi b} with $k$ given different values. \textcolor{black}{The loop structure of the blue line ($v=0.5$), \textcolor{black}{at which comes from particle interaction energy is larger than the optical lattice strength \cite{PhysRevA.86.063636}}, is witnessed around the {edge $k=1$ of the Brillouin zone}. And for the red line ($v=4$) there is no loop structure. Here, the Bloch band also confirms that the static solution in Fig.~\ref{figblochband} are Bloch wave solutions. Actually, the higher value of chemical potential $\mu$, the larger density of total particle $\rho$ in one lattice and hence the larger value of particle interaction energy, which means that when we increase the chemical potential the critical value of the optical lattice strength $v_{c}$ that exceeds the particle interaction energy to prevent the loop structure from appearing will also increase. And we confirm this in Fig.~\ref{fig bloch band2}.}

\section{Instability and sound modes}\label{Landau instability}

\textcolor{black}{Now we study the linear instability of the Bloch wave solutions obtained in the previous section via QNM analyses.} To calculate the QNM, a better choice is to rewrite the static equations of motions \eqref{static Jx}-\eqref{static Psi} with the Bloch wave form for $\psi$ in the infalling Eddington\textcolor{black}{-Finkelstein coordinates} \cite{Li:2020ayr}:
\begin {figure*}
\centering
\begin{subfigure}[b]{0.465\textwidth}
\includegraphics[width = \textwidth]{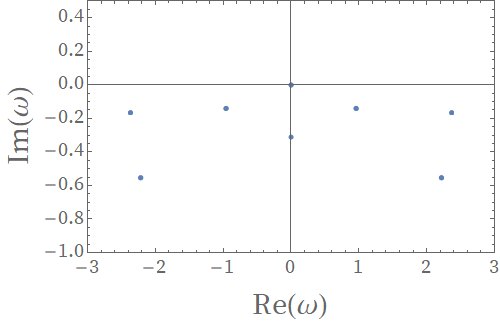}
\caption{QNM with $k=0, q=0$}
\label{figQNMk0q0}
\end{subfigure}
\begin{subfigure}[b]{0.465\textwidth}
\includegraphics[width = \textwidth]{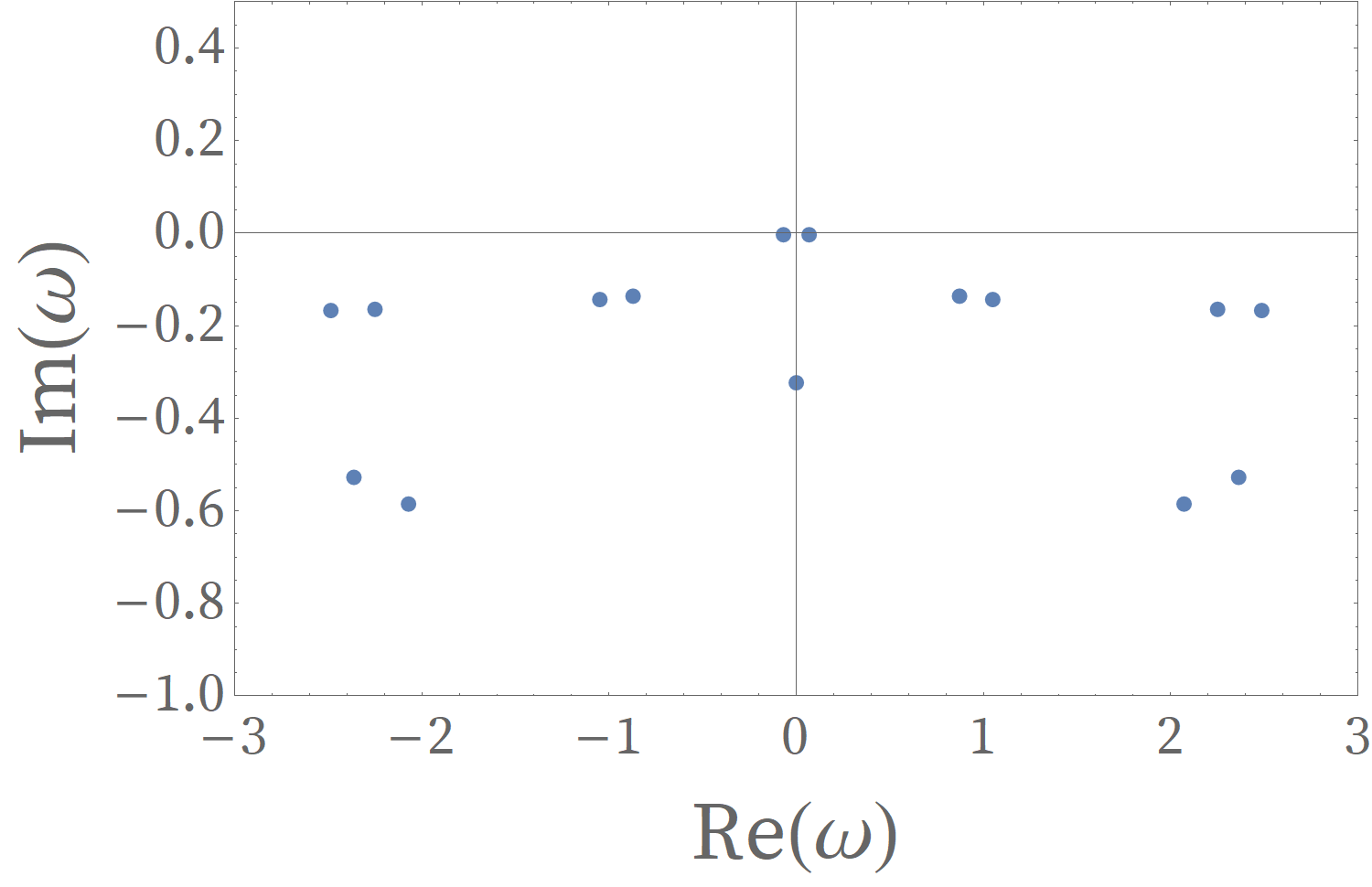}
\caption{QNM with $k=0, q=0.15$}
\label{figQNMk0q015}
\end{subfigure}\\
\begin{subfigure}[b]{0.465\textwidth}
\includegraphics[width = \textwidth]{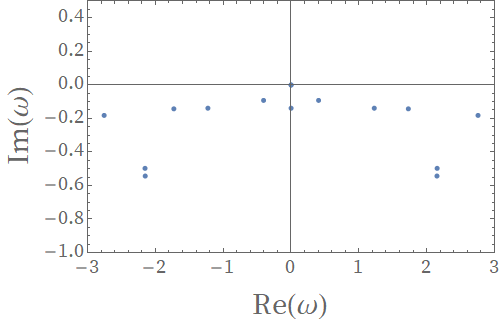}
\caption{QNM with $k=0.75, q=0$}
\label{figQNMk035q0}
\end{subfigure}
\begin{subfigure}[b]{0.465\textwidth}
\includegraphics[width = \textwidth]{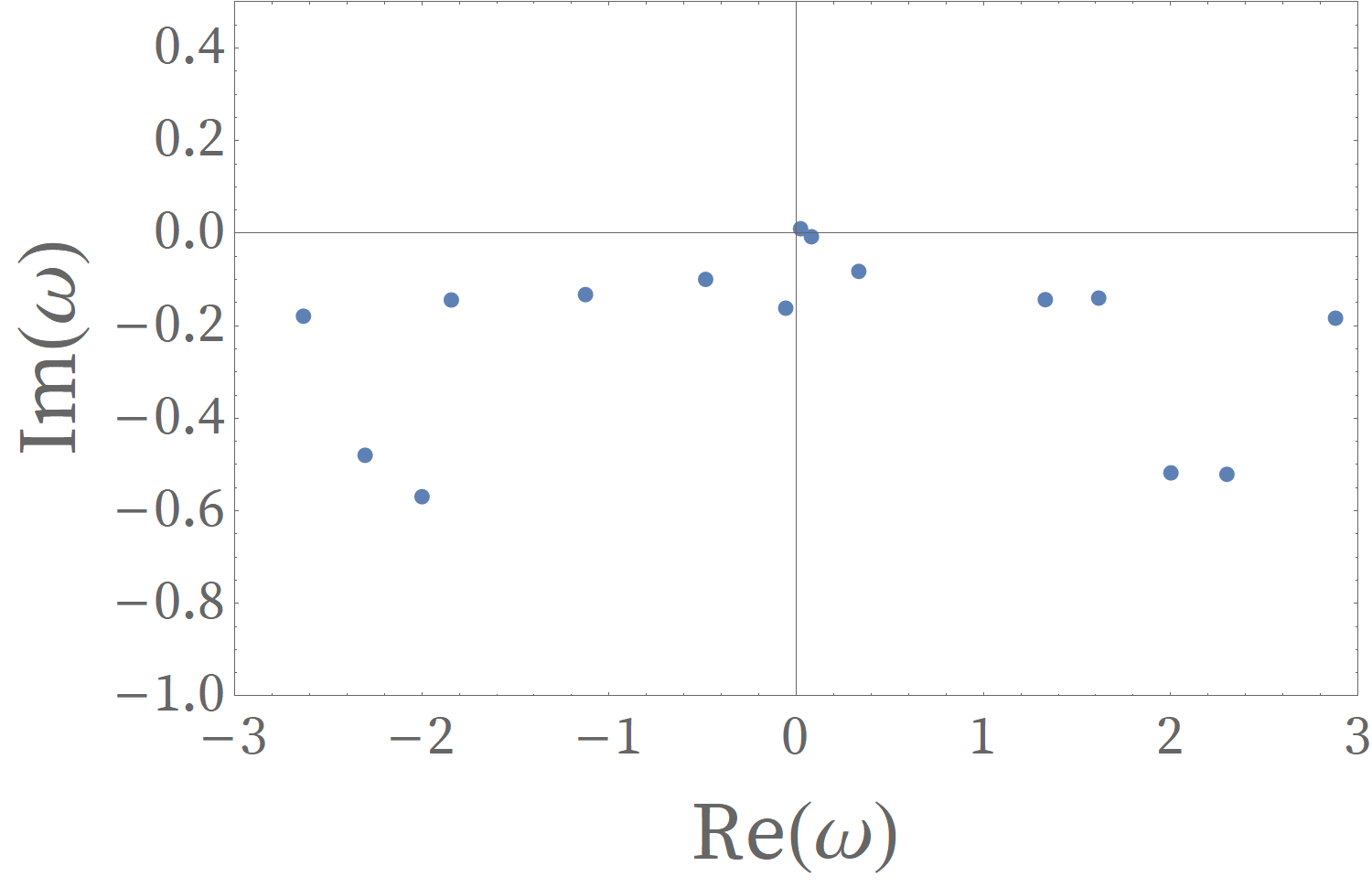}
\caption{QNM with $k=0.75, q=0.15$}
\label{figQNMk035q015}
\end{subfigure}
\caption{\ref{figQNMk0q0} and \ref{figQNMk0q015} show the QNM corresponding to $\psi$'s Bloch wave vector $k=0$ and wave vectors of the perturbed field $q=0$ (left) and $q=0.15$ (right), respectively; \ref{figQNMk035q0} and \ref{figQNMk035q015} show the QNM corresponding to $\psi$'s Bloch wave vector $k=0.75$ and wave vectors of the perturbed field $q=0$ (left) and $q=0.15$ (right), respectively. Actually, the QNM is symmetric under Re($\omega$)$\rightarrow- $Re($\omega$) when q=0 and we confirm this in Appendix \ref{appd:generalized-eigenvalue-problems}.}
\label{figQNM}
\end {figure*}
\begin{equation}\label{Infalling Jx}
2 \partial_{t} \partial_{z}A_{x}-\partial_{z} \partial_{x}A_{t}-\partial_{z}\left(f \partial_{z}A_{x}\right)+i\left(\psi^{*} \partial_{x} \psi-\psi \partial_{x} \psi^{*}-2 i\left(A_{x}-k\right) \psi^{*} \psi\right)=0,
\end{equation}
\begin{equation}\label{Infalling Jz}
\begin{split}
\partial_{t} \partial_{z}A_{t}+\partial_{t} \partial_{x} A_{x}-\partial_{x}^{2} A_{t}-f \partial_{x} \partial_{z}A_{x}+i\left(\psi^{*} \partial_{t} \psi-\psi \partial_{t} \psi^{*}\right)\\
-if\left(\psi^{*} \partial_{z} \psi-\psi \partial_{z} \psi^{*}\right)+2 \psi^{*}A_{t} \psi=0,
\end{split}
\end{equation}
\begin{equation}\label{Infalling Jt}
\left(-\partial_{z}^{2}A_{t}+\partial_{x} \partial_{z} A_{x}\right)+i\left(\psi^{*} \partial_{z} \psi-\psi \partial_{z} \psi^{*}\right)=0,
\end{equation}
\begin{equation}\label{Infalling psi}
\begin{split}
2\partial_{t}\partial_{z}\psi-f \partial_{z}^{2}\psi-\left(f^{\prime}+2iA_{t}\right)\partial_{z}\psi-\partial_{x}^{2}\psi+2 i\left(A_{x}-k\right)\partial_{x}\psi\\
+\left(-i\partial_{z}A_{t}+i\partial_{x}A_{x}+z+\left(A_{x}-k\right)^{2}\right) \psi=0,
\end{split}
\end{equation}
where the metric is
$$ds^{2}=\frac{1}{z^{2}}\left(-f(z) d t^{2}- dtdz+dx^{2}+dy^{2}\right).$$
Here and in the following sections we will not separate $\psi$ into real and imaginary parts. Similarly, there is one constraint equation, \textcolor{black}{which is chosen to be Eq.\eqref{Infalling Jz}}. Next, we give all the bulk field perturbations
\begin{equation}\label{eq:perturbation}
\begin{aligned}
\delta A_{t}&=e^{-i \omega t+i q x} a+e^{i \omega^{*} t-i q x} a^{*}, \\
\delta A_{x}&=e^{-i \omega t+i q x} b+e^{i \omega^{*} t-i q x} b^{*}, \\
\delta \psi&=e^{-i \omega t+i q x} u+e^{i \omega^{*} t-i q x}v^{*}.
\end{aligned}
\end{equation}
Substitute these perturbed fields into the equations of motion, we will get the linear perturbation equations
\begin {figure*}
\begin{subfigure}[b]{0.475\textwidth}
\includegraphics[width = \textwidth]{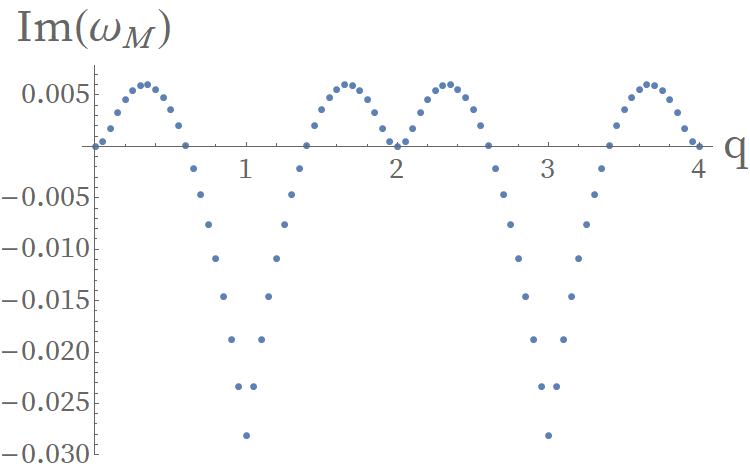}
\caption{$k=0.7, q\in[0, 4]$}
\end{subfigure}
\begin{subfigure}[b]{0.475\textwidth}
\includegraphics[width = \textwidth]{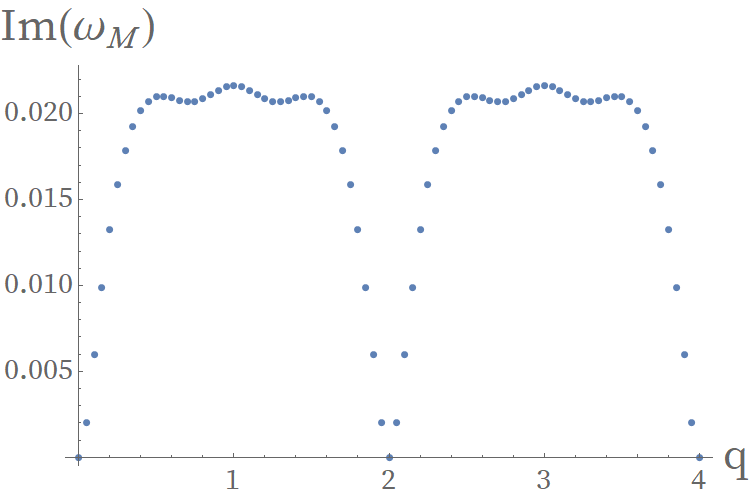}
\caption{$k=0.75, q\in[0, 4]$}
\end{subfigure}
\caption{$\operatorname{Im}(\omega_M)$ as a function of $q$ when Bloch wave vector $k=0.7, 0.75$.}
\label{figomegaq}
\end {figure*}
\begin{equation}\label{eq:perturbationEOMa}
\begin{array}{c}
\partial_{z}\left(\partial_{x} b+i q b\right)+\left(i v \partial_{z} \psi+i \psi^{*} \partial_{z} u\right)-\left(i u \partial_{z} \psi^{*}+i \psi \partial_{z} v\right)-\partial_{z}^{2} a=0,
\end{array}
\end{equation}
\begin{equation}\label{eq:perturbationEOMb}
\begin{array}{c}
\left(2 b \psi \psi^{*}+2\left(A_{x}-k-\frac{1}{2} q\right) \psi^{*}u+2\left(A_{x}-k+\frac{1}{2} q\right) \psi v\right)-2 i \omega \partial_{z} b+ \\
i\left(v \partial_{x} \psi+\psi^{*} \partial_{x} u\right)-i\left(u \partial_{x} \psi^{*}+\psi \partial_{x}v\right)-\partial_{z}\left(f \partial_{z} b\right)-\partial_{z}\left(\partial_{x}a+i q a\right)=0,
\end{array}
\end{equation}
\begin{equation}\label{eq:perturbationEOMu}
\begin{array}{c}
\left(z+\left(A_{x}-k-q\right)^{2}+i\left(\partial_{x} A_{x}-\partial_{z} A_{t}\right)\right) u+\left(2\left(A_{x}-k\right) b+i\left(\partial_{x} b+i q b-\partial_{z} a\right)\right) \psi- \\
\left(2 i \omega+f^{\prime}+2 i A_{t}\right) \partial_{z}u+2i\left(A_{x}-k-q\right) \partial_{x} u-f \partial_{z}^{2} u-\partial_{x}^{2} u-2 i a \partial_{z} \psi+2 ib\partial_{x} \psi=0,
\end{array}
\end{equation}
\begin{equation}\label{eq:perturbationEOMv}
\begin{array}{c}
\left(z+\left(A_{x}-k+q\right)^{2}-i\left(\partial_{x}A_{x}-\partial_{z} A_{t}\right)\right) v+\left(2\left(A_{x}-k\right) b-i\left(\partial_{x}b+iqb-\partial_{z} a\right)\right) \psi^{*}- \\
\left(2 i \omega+f^{\prime}-2i A_{t}\right) \partial_{z} v-2 i\left(A_{x}-k+q\right) \partial_{x}v-f \partial_{z}^{2} v-\partial_{x}^{2} v+2 i a \partial_{z} \psi^{*}-2i b \partial_{x} \psi^{*}=0.
\end{array}
\end{equation}
In \textcolor{black}{accordance with the background steady} solutions, we impose Dirichlet and periodic boundary conditions for $a$, $b$, $u$ and $v$ at $z=0$ and in the $x$ direction, respectively, while \textcolor{black}{regular} boundary conditions are chosen at the horizon $z=1$. Then, we can obtain $\omega$ from the perturbation equations by solving generalized eigenvalue problems  if $k$ and $q$ are given. (See Appendix \ref{appd:generalized-eigenvalue-problems} for more details about this procedure.)

The perturbations in Eq.\eqref{eq:perturbation} will \textcolor{black}{grow exponentially in the linear regime} if there exists an $\omega$ whose imaginary part is positive. We use $\omega_M$ to denote the $\omega$ with the maximal imaginary part when $q$ and $\mu$ are given, so a system is (linearly) unstable if $\operatorname{Im}(\omega_M)>0$. Fig.~\ref{figQNM} shows the result for the distributions of $\omega$ with two different values for Bloch wave vector $k$ and $q$. In Fig.~\ref{figQNMk0q0} and Fig.~\ref{figQNMk0q015}, $\operatorname{Im}(\omega_M)\leq 0$, so the $k=0$ solution is stable under perturbations $q=0$ and $q=0.15$. By comparison, the solution at $k=0.75$ is unstable because $\operatorname{Im}(\omega_M)>0$ in Fig.~\ref{figQNMk035q015}, despite the fact that in Fig.~\ref{figQNMk035q0} $\operatorname{Im}(\omega_M)\leq 0$. A system is stable only when $\operatorname{Im}(\omega_M)\leq 0$ for all values of $q$.

Actually we find that there are two special values $k_{c}$ and $q_{M}$ for the Bloch waves. $k_{c}$ means \textcolor{black}{the critical wave vector exceeding which the Bloch wave vector $k$ will lead to an unstable state}; $q_{M}$ means \textcolor{black}{the Bloch wave vector of the most unstable perturbation mode, which has a maximal imaginary value of $\omega_M$}, since it will increase with time $t$ with the form $e^{\omega_{I}t}$ controlled by the linear perturbation equations in early times (the late time behavior will be more interesting and it will be investigated in the next section.). Fig.~\ref{figomegaq} shows the value of $\omega_{M}$ with respect to $q\in[0, 4]$, the periodicity and $q_{M}$ can be seen directly.

\begin {figure*}
\centering
\begin{subfigure}[b]{0.48\textwidth}
\includegraphics[width = \textwidth]{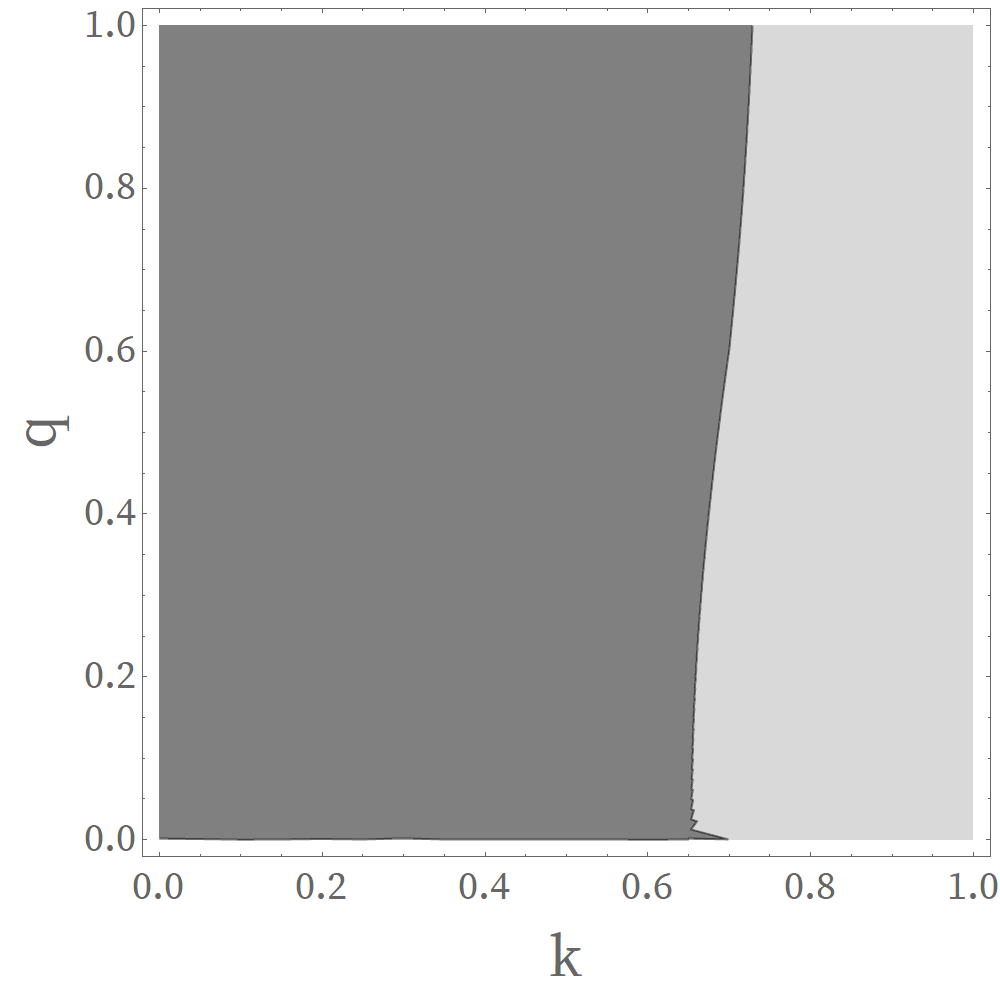}
\end{subfigure}
\caption{Instability diagram for Bloch waves with $k\in[0, 1]$ and $q\in[0, 1]$. \textcolor{black}{In the darker grey region (the left part in the figure), $\operatorname{Im}(\omega_M)\leq 0$, and in the lighter grey region (the right part in the figure), $\operatorname{Im}(\omega_M)>0$.}}
\label{instability}
\end {figure*}

As we have explained, when $q$ and $k$ are given, $\omega_M$ can be obtained to determine the stability under the corresponding perturbation. In this spirit, we plot stable and unstable regions as a function of $q$ and $k$, i.e. the instability diagram, in \textcolor{black}{the half first Brillouin zone} in Fig.~\ref{instability}. In this plot, the left region in the figure, which is colored darker grey, is stable ($\operatorname{Im}(\omega_M)\leq 0$), and the right region in the figure, which is colored lighter grey, is unstable ($\operatorname{Im}(\omega_M)> 0$).\footnote{When $q=0$, $\operatorname{Im}(\omega_M)\leq 0$ for all $k$, which cannot be shown clearly in the figure.} The figure also tells us that in the unstable region the cutoff $q$ increases with $k$, and when $k$ is large enough (but still in the first Brillouin zone), the system is unstable for all values of $q$ \textcolor{black}{except $q=0$}.

\begin{figure*}
\centering
\begin{subfigure}[b]{0.45\textwidth}
\includegraphics[width = \textwidth]{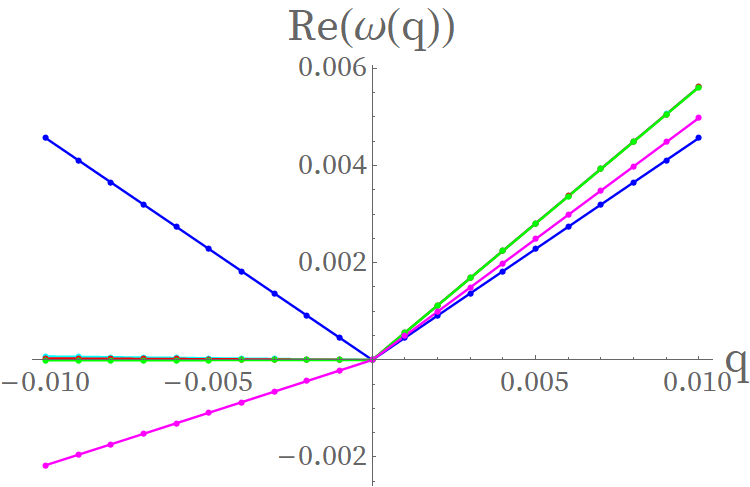}
\end{subfigure}
\begin{subfigure}[b]{0.45\textwidth}
\includegraphics[width = \textwidth]{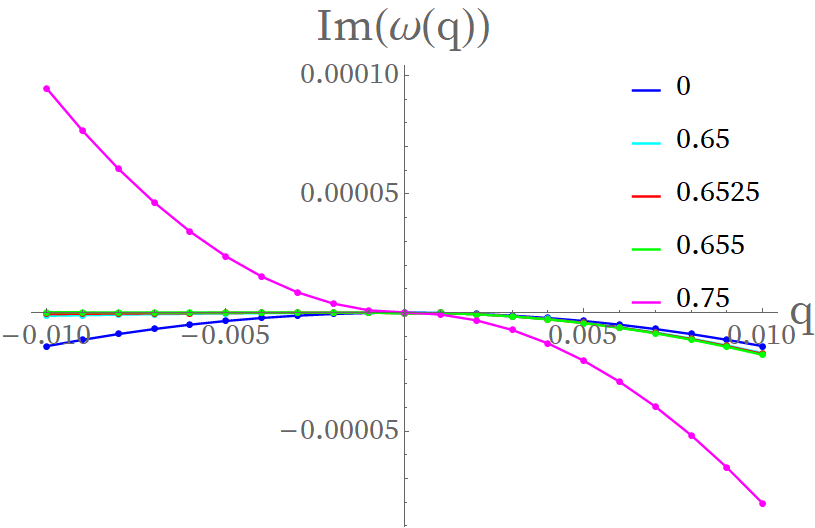}
\end{subfigure}
\caption{Dispersion relations for sound modes at $k=0, 0.65, 0.6525, 0.655, 0.75$. Left panel shows $\operatorname{Re}(\omega(q))$ and right panel shows $\operatorname{Im}(\omega(q))$. A universal phenomenon is that $\operatorname{Im}(\omega(q))>0$ always accompanies $\operatorname{Re}(\omega(q))<0$. }
\label{sound}
\end {figure*}

As a signal of onset of the instability mentioned in \cite{2014JHEP...02..063A,2020arXiv201006232L}, we also plot the dispersion relation of the sound mode in Fig.\ref{sound}. When there is a non-zero $k$ the sound speed becomes direction-dependent. There are two directions corresponding to the maximal and minimal values of the sound speed, which are parallel ($q>0$) and anti-parallel ($q<0$) to the velocity of the superflow, respectively. Hereafter we denote \textcolor{black}{the sound speed} as $c_{+}$ for $q>0$ and $c_{-}$ for $q<0$.

The onset of instability is accompanied by the sign-change of $c_{-}$. Along the negative axis of $q$, $\operatorname{Re}(\omega)$ becomes negative as $k$ is larger than 0.655. As $\operatorname{Re}(\omega)$ corresponds to the energy of the perturbation, $c_{-}<0$ (so $\operatorname{Re}(\omega)<0$) indicates the existence of \textcolor{black}{a perturbed} state with lower energy and thus the instability of the system (Landau instability). At the same time, we find that $\operatorname{Im}(\omega)>0$, which implies the dynamical instability of the system. So, our result confirms that the Landau instability and dynamical instability occur at the same time, as we have mentioned.

For the purpose of opening the next section we give a brief discussion about this section. Since it is not always true for a superflow to flow in the optical lattice stably, when $k\ge k_{c}$ \textcolor{black}{excitations will always appear to slow the superflow down \cite{2020arXiv201006232L}}, but what kind of intermediate states will this unstable superflow go through and what kind of final states will it end up with? In the next section we will use full nonlinear evolution to find the final state out for the Bloch wave-type superflow and the intermediate states will also be discussed.

\section{Real time evolution for stable and unstable superflows}\label{Real time evolution}

To simulate the evolution of the holographic superfluid, we adopt the same scheme as in \cite{Li:2020ayr}, i.e. taking Eqs.\eqref{Infalling Jx}, \eqref{Infalling Jt} and \eqref{Infalling psi} as evolution equations while choosing Eq.\eqref{Infalling Jz} as the constraint. Since Eq.\eqref{Infalling Jt} does not have time derivative terms but have space derivatives up to the second order, it can be solved with one boundary condition from $A_{t}(z_{b})=4.5+0.5\cos(2x)$ and another from the restriction of Eq.\eqref{Infalling Jz} on the conformal boundary. In the spatial directions  the pseudo-spectral method is chosen for discretization, while in the time direction we adopt the fourth order Runge-Kutta method. When doing the time evolution, we choose static solutions, including $\psi$, $A_{t}$ and $A_{x}$ which are solved from the static equations of motion, as the initial states (at $t=0$), and at an early time $t_{1}$ we add perturbations, which are some random functions with small amplitudes, to $\psi$. Here, we require that the perturbations should not break the source free boundary condition of $\psi$. The perturbation is not necessarily periodic and there are many \textcolor{black}{lattice cells} in experiments, so it is necessary to include more \textcolor{black}{lattice cells} into the time evolution.\footnote{Actually a larger number of lattices for the simulation is more realistic and reliable, but in practice we choose $11$ \textcolor{black}{cells}, which is enough to study some universal properties in our case.}

\subsection{Evolution of particle current density}

\begin {figure*}
\begin{subfigure}[b]{0.3\textwidth}
\includegraphics[width = \textwidth]{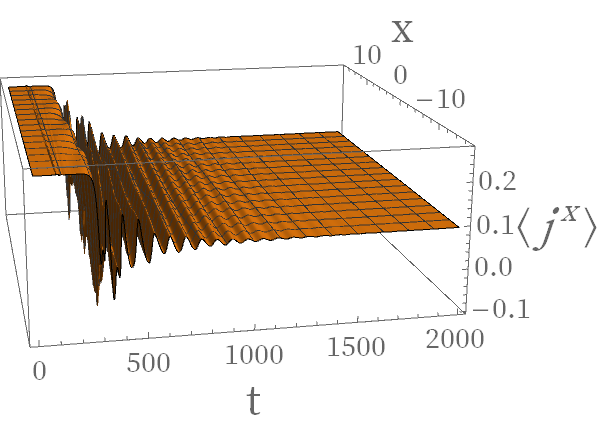}
\caption{$\langle j^{x}(t,x)\rangle$}
\label{figrhok04t0}
\end{subfigure}
\begin{subfigure}[b]{0.3\textwidth}
\includegraphics[width = \textwidth]{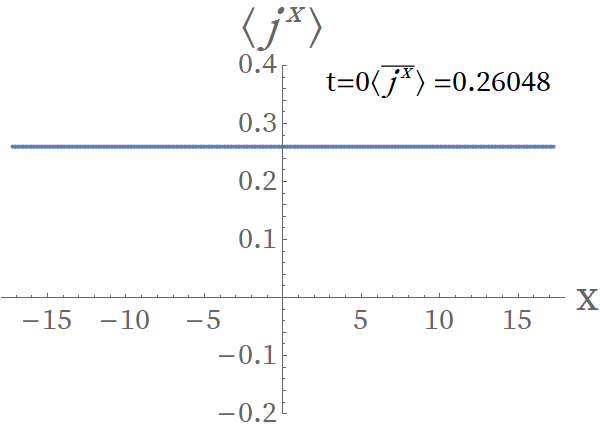}
\caption{$t=0$}
\label{figrhok04t0}
\end{subfigure}
\begin{subfigure}[b]{0.3\textwidth}
\includegraphics[width = \textwidth]{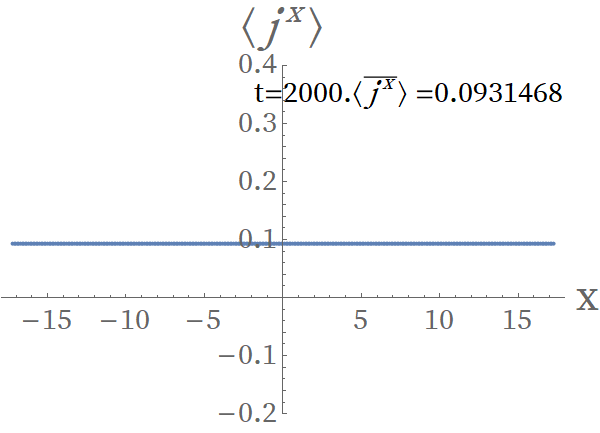}
\caption{$t=2000$}
\label{figrhok04t15125}
\end{subfigure}\\
\begin{subfigure}[b]{0.3\textwidth}
\includegraphics[width = \textwidth]{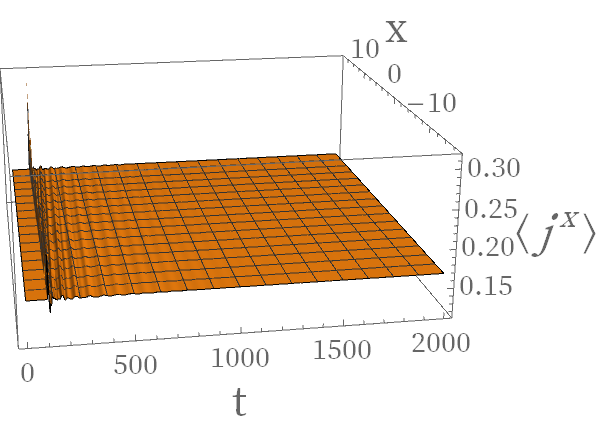}
\caption{$\langle j^{x}(t,x)\rangle$}
\label{figrhok01t0}
\end{subfigure}
\begin{subfigure}[b]{0.3\textwidth}
\includegraphics[width = \textwidth]{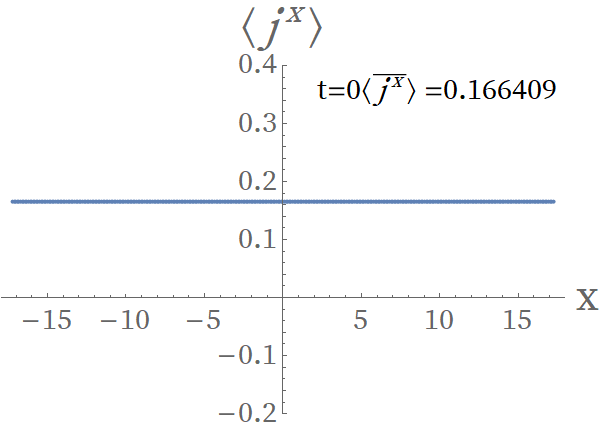}
\caption{$t=0$}
\label{figrhok01t0}
\end{subfigure}
\begin{subfigure}[b]{0.3\textwidth}
\includegraphics[width = \textwidth]{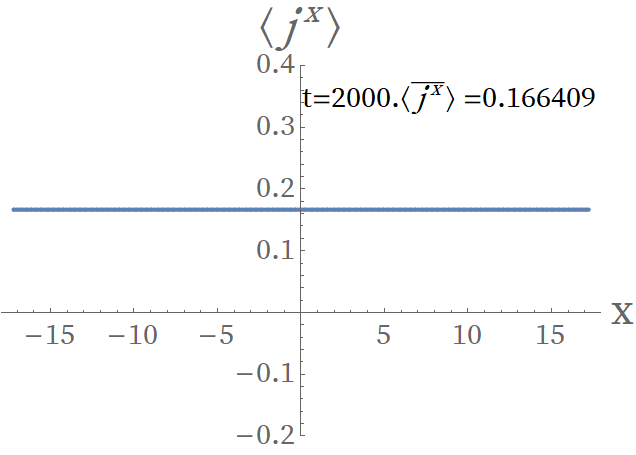}
\caption{$t=2000$}
\label{figrhok01t15125}
\end{subfigure}
\caption{Evolutions for superflows with $k=0.3$ (stable) and $k=0.75$ (unstable), respectively. These panels show the particle current density with $t\in[0,2000]$.}
\label{evolrho}
\end {figure*}
\textcolor{black}{We introduce perturbations at $t=100$, i.e. $t_1=100$, and observe the evolution of the systems with initial Bloch vectors $k$ chosen as $0.3$ and $0.75$, respectively. Fig.~\ref{evolrho} shows the evolution of the \textcolor{black}{particle current density} $\langle j^{x}\rangle$, which is defined as
\begin{equation}
    \langle j^{x}\rangle=\lim_{z\to 0}\sqrt{-g}F^{zx}.
\end{equation}}

\textcolor{black}{From Fig.~\ref{evolrho}, the evolution of $\langle j^{x}\rangle$ of an unstable superflow can be divided into five stages. In the first stage $t<100$, no change is observed, which confirms that the static solutions obtained in the Schwarzschild coordinates is correctly transformed to that in the \textcolor{black}{Eddington-Finkelstein coordinates}. When $100<t<200$, the current density changes slightly since the influence of the perturbation is small; when $200<t<400$, the current density becomes chaotic, which comes from the \textcolor{black}{nonlinear development} of the instability. When $400<t<1500$, the chaotic behavior of the current density disappears but the current is still inhomogeneous; In this stage, the system \textcolor{black}{is tending to a steady} state. In the last stage $t>1500$, the system becomes steady with the final current density $\langle j^{x}\rangle_{f}=0.0931468$, which is much smaller than the initial value 0.26048. In contrast, for the stable superflow ($k=0.3$ as an example for comparison) the situation is totally different. The influence of perturbations on the system is insignificant and the system will \textcolor{black}{evolve to a state that is same as the initial state in a few moments.}}

\subsection{Evolution of condensate and intermediate states}
\begin {figure*}
\centering
\begin{subfigure}[b]{0.4\textwidth}
\includegraphics[width = \textwidth]{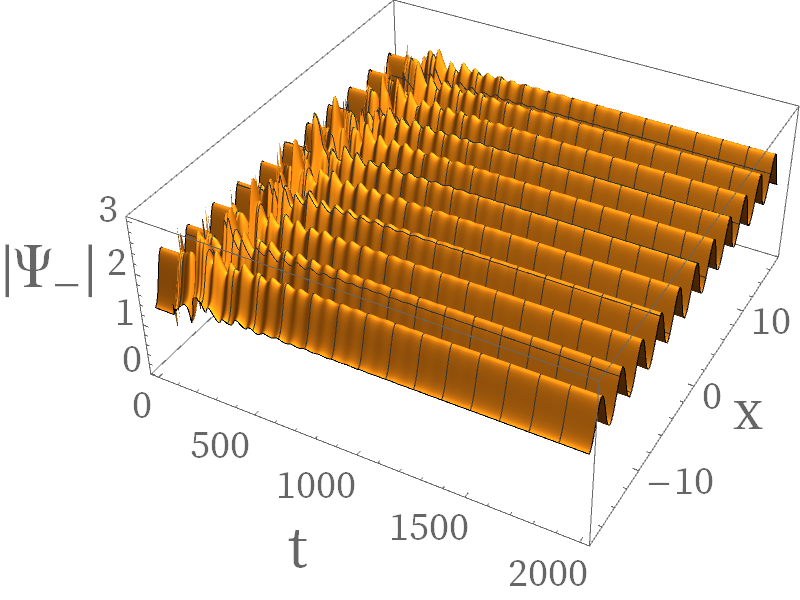}
\caption{$|\Psi_{-}|(t,x)$}
\label{figrsolitont1000}
\end{subfigure}
\begin{subfigure}[b]{0.4\textwidth}
\includegraphics[width = \textwidth]{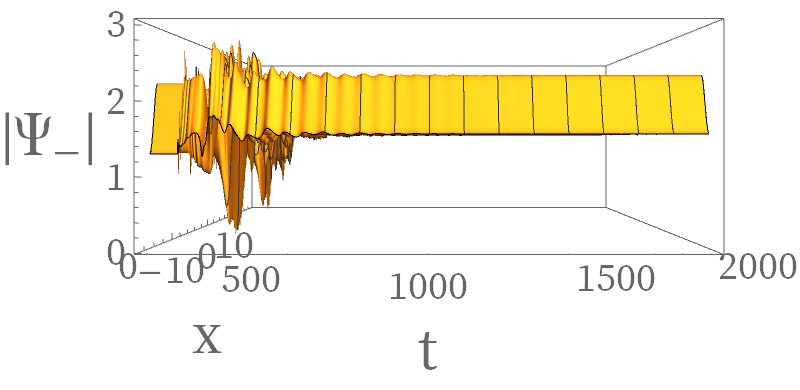}
\caption{$|\Psi_{-}|(t,x)$}
\label{figrsolitont500}
\end{subfigure}\\
\begin{subfigure}[b]{0.355\textwidth}
\includegraphics[width = \textwidth]{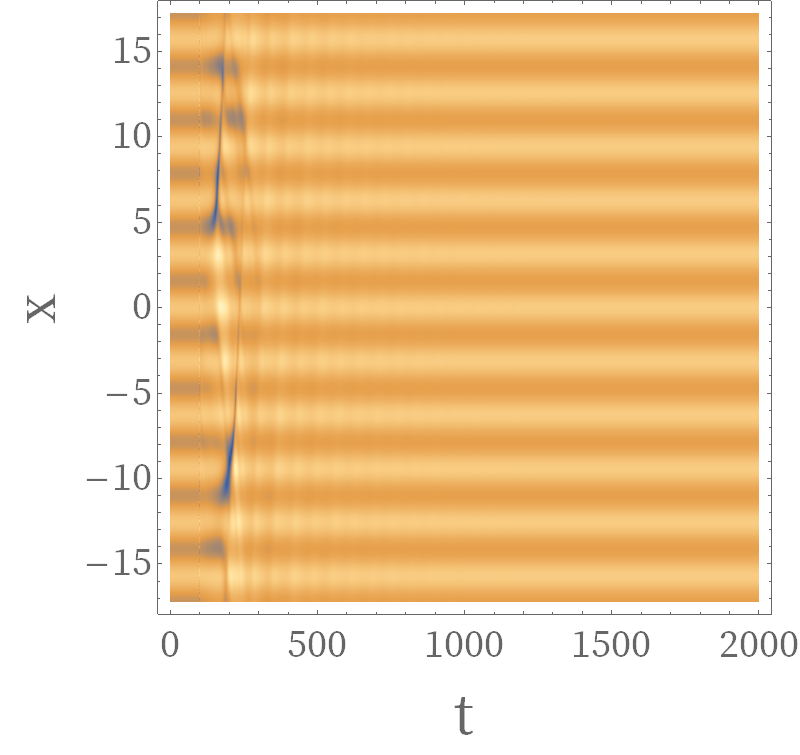}
\caption{$|\Psi_{-}|(t,x)$}
\label{figrsolitont0}
\end{subfigure}
\begin{subfigure}[b]{0.04\textwidth}
\includegraphics[width = \textwidth]{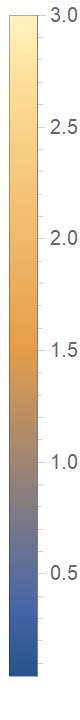}
\label{figrsolitont50}
\end{subfigure}
\begin{subfigure}[b]{0.4\textwidth}
\includegraphics[width = \textwidth]{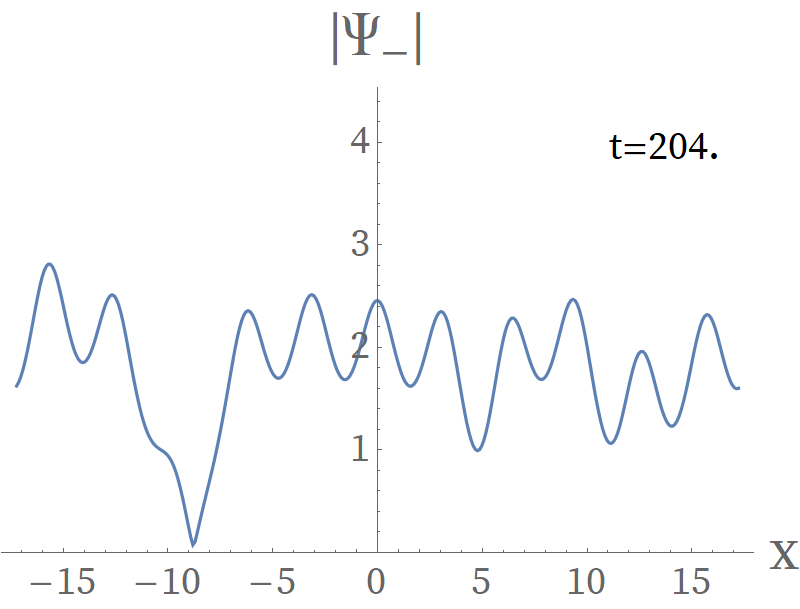}
\caption{$t=204$}
\label{figrsolitont7432}
\end{subfigure}
\caption{Evolution of the condensate $|\Psi_{-}(t,x)|$ with $t\in[0,2000]$ of the unstable superflow $k=0.75$. The perturbation is given at $t=100$; After that there are some solitons generated, as plotted in Fig.~\ref{figrsolitont7432}.}
\label{evolsoliton}
\end {figure*}
\textcolor{black}
{As we have seen from the evolution of $\langle j^{x}\rangle$, the intermediate states of the system are chaotic and complicated. \textcolor{black}{However,} there are some universal features which can be extracted from the evolution of the condensate $|\Psi_{-}|$ and we plot them in Fig.~\ref{evolsoliton}.}

\textcolor{black}{We plot the evolution of $|\Psi_{-}|$ from different viewing angles, i.e. Fig.~\ref{figrsolitont1000}, Fig.~\ref{figrsolitont500} and Fig.~\ref{figrsolitont0}. Fig.~\ref{figrsolitont1000} is from an ordinary viewing angle, from which we can see the complete picture of $|\Psi_{-}(t,x)|$; Fig.~\ref{figrsolitont500} is plotted with the viewing angle along the $x$ axis. We can see that the condensate during $100<t<400$ is also chaotic and the value of $|\Psi_{-}|$ at later times is larger than its initial value. The planform of $|\Psi_{-}|$ is plotted at Fig.~\ref{figrsolitont0} and \textcolor{black}{nodes of the order parameter} appear during $100<t<400$, which indicates the formation of solitons. We select one \textcolor{black}{moment} ($t=204$) of the soliton formation in Fig.~\ref{figrsolitont7432}. The formation of solitons is natural here \cite{2020arXiv201006232L} since the unstable superflow is thought of as having a higher free energy than a less unstable superflow with soliton excitations. \textcolor{black}{Along with the dissipative processes, the whole system will go to a stable} state with a lower free energy.}

\subsection{Final stable state}
Since both $\langle j^{x}\rangle$ and $|\Psi_{-}|$ become time independent at the end of the evolution, the final state of the system must be able to be described as a Bloch state just like the initial steady flow states that are solved in Sec.~\ref{Bloch waves}. This is indeed the case. Actually, the final state can also be solved from the equations of motion, i.e. \eqref{Jx}, \eqref{Jt}, \eqref{Psi a} and \eqref{Psi b}, with an additional boundary condition besides the boundary conditions given in Sec.~\ref{Bloch waves}, while not specifying the Bloch wave vector $k$.

\begin{figure*}
\centering
\begin{subfigure}[b]{0.45\textwidth}
\includegraphics[width = \textwidth]{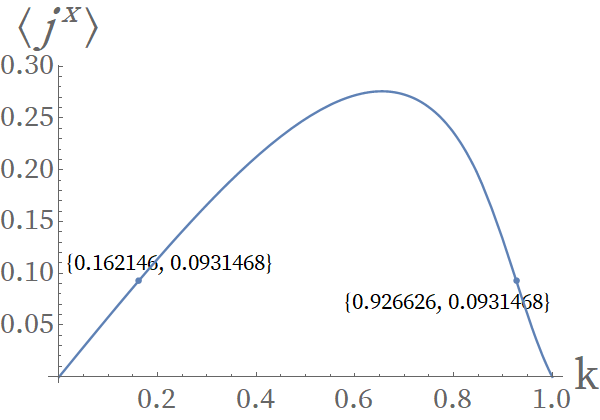}
\caption{$\langle j^{x}(k)\rangle$}
\label{jk}
\end{subfigure}
\begin{subfigure}[b]{0.45\textwidth}
\includegraphics[width = \textwidth]{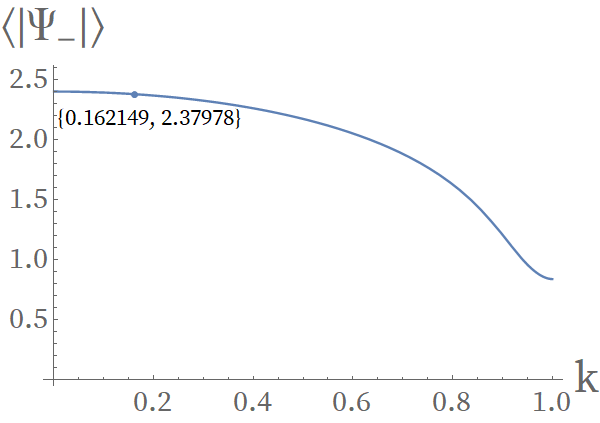}
\caption{$\langle|\Psi_{-}|(k)\rangle$}
\label{conk}
\end{subfigure}
\caption{Relations between $\langle j^{x}\rangle$ and $k$ as well as $\langle|\Psi_{-}|\rangle$ and $k$. Both figures are obtained by interpolation with discrete value of k, so the value 0.162146, 0.926626 and 2.37978 is not exact. \textcolor{black}{The non-monotonicity for $\langle j^{x}\rangle$ gives two states with different $k$, while the monotonicity for $\langle|\Psi_{-}|\rangle$ excludes the larger $k$ state.}}
\label{jandcon}
\end {figure*}
\textcolor{black}{When there is a one-to-one correspondence between $\langle j^{x}\rangle$ and the Bloch wave vector $k$, one can get the additional boundary condition directly by fixing $\langle j^{x}\rangle=\langle j^{x}\rangle_{f}$. However, as we can find from Fig.~\ref{jk}, i.e. the relation between $\langle j^{x}\rangle$ and $k$, a fixed $\langle j^{x}\rangle$ normally corresponds to two values of $k$. To make the solution unique, we can take account of the average condensate $\langle|\Psi_{-}|\rangle$ in one lattice cell. From Fig.~\ref{figrsolitont1000} we know that the condensate amplitude $|\Psi_{-}|$ of a steady state is not a constant, so the relation between $|\Psi_{-}|$ and $k$ is not clear, while $\langle|\Psi_{-}|\rangle$ is a monotonically decreasing function of $k$, as plotted in Fig.~\ref{conk}. Therefore, the boundary condition can be uniquely determined by combining the two considerations:  $\langle j^{x}\rangle=\langle j^{x}\rangle_{f}$ and the value of $\langle|\Psi_{-}|\rangle$. From Fig.~\ref{figrsolitont500} and Fig.~\ref{figrhok04t15125} we conclude that the final Bloch state corresponds to a larger $\langle|\Psi_{-}|\rangle$ with $\langle j^{x}\rangle=0.0931468$, thus we can solve the equations of motion with the additional boundary condition $\partial_{z}{A_x}|_{z=0}=0.0931468$, and the solution is marked on Fig.~\ref{conk}.}

\section{Summary}\label{Summary}

The dynamical process of superfluids that simultaneously experiences Landau and dynamical instability in optical lattice is investigated using the simplest holographic superfluid model. Due to the existence of dissipation the sound mode of an unstable superflow will always have negative energy and grows up exponentially at early times. We give a universal picture of the instability process for the unstable superflow. From real time evolution the unstable superflow is shown to reduce its particle current density to settle down to a steady state with a current density $\langle j^{x}\rangle_{f}$ and a Bloch wave vector $k_{f}$ below the critical value $k_{c}$, where this $k_{f}$ can be solved from the equations of motion by fixing $\langle j^{x}\rangle=\langle j^{x}\rangle_{f}$. \textcolor{black}{In the course of the evolution, a chaotic process involving soliton generation is observed, in which the solitons are supposed to play the role of reducing the wave vector $k$ (similar to the case without an optical lattice \cite{2020arXiv201006232L}).}

\textcolor{black}{As we have mentioned, when the number of dimensions of the boundary system is greater than one, the process for an unstable state to evolve into a stable one will become much more complicated, which to some extent is described as transient turbulence \cite{2020arXiv201006232L}. So beside the one dimensional boundary superfluid system in optical lattice, it is worthwhile to consider higher dimensional cases. Another direction for future investigation is to include backreaction in this holographic superfluid model, which will enable us to explore the complete interplay between the normal fluid and superfluid components of the superflow at a finite temperature in optical lattice.}

\begin{acknowledgments}
\textcolor{black}{We are grateful to Biao Wu and Hongbao Zhang for their helpful discussions. Peng Yang is also grateful to Shanquan Lan for his helpful discussions. Xin Li acknowledges the support form China Scholarship Council (CSC, No.~202008610238). This work is partially supported by NSFC with Grant No.11975235 and 12035016.}
\end{acknowledgments}

\appendix
\section{{Notes on generalized eigenvalue problems}}\label{appd:generalized-eigenvalue-problems}

Let's begin at the bulk fields perturbation \eqref{eq:perturbation}. It's well known that when a system has the symmetry of time translation, then any field $F(t,x)$ can be separated into mode $f(x)e^{-i\omega t}$ with different $\omega$. The situation is
similar for space. If there is a continuous space translation symmetry, then we have $F(t,x)\sim f(t)e^{ikx}$ with continuous value of $k$; if there is a discrete space translation symmetry, then we also have $F(t,x)\sim f(t)e^{ikx}$ but with discrete value of $k=\frac{2\pi}{L}n$, where $n$ is integer and $L$ is lattice constant.

But in our case, we do not use the above separation though there is a discrete space translation symmetry but use Bloch wave to describe the $x$ direction of the fields, i.e. $F(t,x)=f(t,x)e^{iqx}$. Here F(t,x) and f(t,x) can be complex functions and we can do some deformations,
\begin{equation}
\begin{split}
F(t,x)&=f(t,x)e^{iqx}\\
&=\frac{1}{2}(f(t,x)e^{iqx}+f^{*}(t,x)e^{-iqx})+\frac{1}{2}(f(t,x)e^{iqx}-f^{*}(t,x)e^{-iqx})\\
&=\frac{1}{2}(f(t,x)+f^{*}(t,x)e^{-2iqx})e^{iqx}+\frac{1}{2}(f(t,x)e^{2iqx}-f^{*}(t,x))e^{-iqx}\\
&=g(t,x)e^{iqx}+h(t,x)^{*}e^{-iqx}.
\end{split}
\end{equation}
With this deformation we can write \eqref{eq:perturbation} and get \eqref{eq:perturbationEOMa}-\eqref{eq:perturbationEOMv} straightforward.

Equations \eqref{eq:perturbationEOMa}-\eqref{eq:perturbationEOMv} can be written as a matrix form, i.e.,
\begin{equation}\label{R}
R\left(\begin{array}{l}
a \\
b \\
u \\
v
\end{array}\right)=\omega A\left(\begin{array}{l}
a \\
b \\
u \\
v
\end{array}\right),
\end{equation}
with appropriate boundary conditions given in Section \ref{Landau instability}. These function is indeed the general eigenvalue function and there are many methods in textbook to solve this function to get the eigenvalue $\omega$.

Before we calculate the value of $\omega$ we can know that if $\omega_{R}$ is real part of one of the eigenvalue $\omega$, then there must be another value of $\omega$ whose real part equals to -$\omega_{R}$. To show this we can define a matrix $L$, which is defined as
\begin{equation}
L=R-\omega A,
\end{equation}
then Eq.\eqref{R} can be written as
\begin{equation}
L\left(\begin{array}{l}
a \\
b \\
u \\
v
\end{array}\right)=0.
\end{equation}
Separate $\omega$ into real part and imaginary pary and we get that the imaginary part of $L$ only contains $\omega_{R}$, which means
\begin{equation}
L_{I}(\omega_{R})\left(\begin{array}{l}
a \\
b \\
u \\
v
\end{array}\right)=0.
\end{equation}
From Equations \eqref{eq:perturbationEOMa}-\eqref{eq:perturbationEOMv} it is easy to get that when $q=0$ the following relation exists, i.e.,
\begin{equation}\label{L}
L_{I}(\omega_{R})\left(\begin{array}{l}
a \\
b \\
u \\
v
\end{array}\right)=-L_{I}(-\omega_{R})\left(\begin{array}{l}
a \\
b \\
v \\
u
\end{array}\right).
\end{equation}
Since the boundary conditions for $u$ and $v$ are the same, changing the places between $u$ and $v$ do not affect the eigenvalues. While when $q\neq0$, there do not have the relation \eqref{L}.
\bibliographystyle{unsrt}
\bibliography{LatexHSFinOL}

\end{document}